\documentclass[11pt]{article}
\usepackage{authblk,fullpage,setspace}
\makeatletter
\newcommand{\printfnsymbol}[1]{\textsuperscript{\@fnsymbol{#1}}}
\makeatother

\begin{document}
\renewcommand*{\Authfont}{\large}
\renewcommand*{\Affilfont}{\normalsize}
\onehalfspacing

\title{Matching reads to many genomes with the $r$-index}
\author[1]{Taher Mun\thanks{Corresponding author: {\tt tmun1@jhu.edu}}}
\author[2]{Alan Kuhnle} 
\author[2]{Christina Boucher\thanks{Equal contribution, ordered alphabetically.}}
\author[3]{\\Travis Gagie\printfnsymbol{2}}
\author[1]{Ben Langmead\printfnsymbol{2}}
\author[4]{Giovanni Manzini\printfnsymbol{2}}

\affil[1]{Department of Computer Science,

Johns Hopkins University, Baltimore, MD \bigskip}

\affil[2]{Department of Computer and Information Science and Engineering,

University of Florida, Gainesville, FL \bigskip}

\affil[3]{Faculty of Computer Science, Dalhousie University, Halifax, Canada

Center for Biotechnology and Bioengineering, Santiago, Chile

School of Computer Science and Telecommunications,

Universidad Diego Portales, Santiago, Chile \bigskip}

\affil[4]{Department of Science and Technological Innovation,

University of Eastern Piedmont, Alessandria, Italy}
\maketitle

\begin{abstract}
The $r$-index is a tool for compressed indexing of genomic databases for exact pattern matching, which can be used to completely align reads that perfectly match some part of a genome in the database or to find seeds for reads that do not.  This paper shows how to download and install the programs {\tt ri-buildfasta} and {\tt ri-align}\,; how to call {\tt ri-buildfasta} on a FASTA file to build an $r$-index for that file; and how to query that index with {\tt ri-align}\,.

{\bf Availability: } The source code for these programs is released under GPLv3 and available at {\tt https://github.com/alshai/r-index}\,.
 \end{abstract}

\section{Background}
\label{sec:background}

The Burrows-Wheeler Transform (BWT)~\cite{BW94} and FM-index~\cite{FM05} are central to the most popular short-read aligners, such as BWA~\cite{LD09} and Bowtie~\cite{LTPS09,LS12}, but until recently it was not known how to apply these concepts effectively to whole genomic databases.  Building on previous authors' work~\cite{MNSV10}, Gagie, Navarro and Prezza~\cite{GNP18} described how a fully functional variant of the FM-index for such a database could be stored in reasonable space: their variant takes $O (r)$ machine words, where $r$ is the number of runs in the BWT of the database, and thus is called the {\em $r$-index}.  Prezza~\cite{Pre18} gave a preliminary implementation, which was significantly extended by Boucher et al.~\cite{BGKLMM19} and Kuhnle et al.~\cite{KMBGLM19}.  This paper is meant as a brief guide to the extended implementation.  For help troubleshooting or to provide feedback, please submit an issue to our GitHub page, which also has more documentation.

\section{Installation}
\label{sec:installation}

In this section we give installation instructions, assuming users are using Ubuntu with \texttt{c++17}-compliant compilers and are familiar with the Unix command line.   If they are using another Unix-like system then they should substitute the appropriate package manager for {\tt apt}\,.

Users should first download some prerequisite packages, and the source code from the github repository:
\begin{verbatim}
$   apt-get update
$   apt-get install -y build-essential cmake git python3 zlib1g-dev
$   git clone --recursive https://github.com/alshai/r-index
\end{verbatim}
They should then compile the code, as follows; any missing dependencies will be automatically downloaded and compiled locally.
\begin{verbatim}
$   cd r-index
$   mkdir build
$   cd build
$   cmake ..
$   make
$   make install
$   cd ..
\end{verbatim}

These commands will install the binaries {\tt ri-buildfasta} and {\tt ri-align} in the system's default \texttt{bin} location (e.g., \texttt{/usr/local/bin} for Ubuntu users), together with {\tt bigbwt}~\cite{BGKLMM19} and the {\tt SDSL} library~\cite{GBM14} (if it is not already present).  If users want the binaries elsewhere, then they should use
\begin{verbatim}
$   cmake -DCMAKE_INSTALL_PREFIX=<dest> ..
\end{verbatim}
instead of \enskip {\tt cmake ..} \enskip in the sequence of commands above, where \texttt{<dest>} is the desired destination directory.

\section{Construction}
\label{sec:construction}

Users can call \texttt{ri-buildfasta} to build the $r$-index for a collection of genomic sequences stored in a FASTA file, which can be gzipped.  For example, to index the 2.7 MB file {\tt data/dengue/genome.fa.gz}, which contains 2042 Dengue Type 1 genomes downloaded from the Virus Pathogen Database and Analysis Resource (ViPR)~\cite{PSZ+12} and occupies 22 MB uncompressed, they can use the following command: 
\begin{verbatim}
$   ri-buildfasta -o dengue data/dengue/genome.fa.gz
\end{verbatim}

This command saves an $r$-index as the two files \texttt{dengue.ri}\,, which contains the main data structures for the index, and \texttt{dengue.1.ri}\,, which contains mappings from offsets in the indexed text to the names of the sequences starting at those offsets.  If users want the {\tt .ri} files to have different names, they should change the \texttt{-o} parameter in the command.  In this example, the {\tt .ri} files have total size 2.4 MB, or 11\% and 89\% of the sizes of the indexed text when uncompressed and gzipped, respectively; once they have been built, {\tt data/dengue/genome.fa.gz} is no longer necessary.

By default, {\tt ri-buildfasta} uses \texttt{bigbwt}, which efficiently handles most large repetitive text collections.  For very small or insufficiently repetitive collections, however, the \texttt{sais} algorithm~\cite{NZC10} might be faster, so users can change the construction algorithm using the \texttt{-b} parameter, as follows:
\begin{verbatim}
$   ri-buildfasta -b sais -o dengue data/dengue/genome.fa.gz
\end{verbatim}

On an Intel i7-7700HQ 2.80 GHz laptop, building the index for {\tt data/dengue/genome.fa.gz} with {\tt bigbwt} takes about 3 seconds and 47 MB of memory while building it with {\tt sais} takes about 10 seconds and 104 MB of memory.  With {\tt bigbwt} we have built a 665 MB $r$-index for a collection of 2000 copies of human chromosome 19 from different individuals, which take 110 GB uncompressed, in under 5 hours on a server using 41 GB of RAM and one thread.  At the time of writing the use of multiple processors does not significantly reduce the construction time, but that should change in the near future.

\section{Alignment}
\label{sec:alignment}

Users can call {\tt ri-align} to search for a collection of patterns stored in a FASTQ file.  For example, to use the $r$-index from the previous section to count the occurrences of the 1000 simulated 100-bp reads in the 219 KB file {\tt data/dengue/reads.fq}, which were taken from the first genome in the dengue collection, they should use the following command:
\begin{verbatim}
$   ri-align count dengue data/dengue/reads.fq
\end{verbatim}
Note that the index should be specified without the \texttt{.ri} extension.  The output is piped to the standard output by default, but can of course be redirected to a file.  If users actually want the positions of the reads' occurrences, they can call {\tt ri-align locate}\,:
\begin{verbatim}
$   ri-align locate dengue data/dengue/reads.fq
\end{verbatim}

By default, {\tt ri-align locate} returns the positions of all the occurrences of each read in the indexed text.  This can be very slow, especially if the read occurs many times, so users can include \texttt{--max-hits <k>} to limit the number of occurrences that are reported, where {\tt <k>} is the desired number of occurrences:
\begin{verbatim}
$   ri-align --max-hits <k> locate dengue data/dengue/reads.fq
\end{verbatim}
Users can include \texttt{--max-range <k>} to obtain the occurrences only of reads that occur at most {\tt <k>} times:
\begin{verbatim}
$   ri-align --max-range <k> locate dengue data/dengue/reads.fq
\end{verbatim}

On the same Intel i7-7700HQ 2.80 GHz laptop and with the output redirected to a file, counting the occurrences of each read takes a total of about 0.06 seconds and locating one occurrence of each read takes a total of about 0.1 seconds.

\section{Interpreting Counts}
\label{sec:counts}

The output of the \texttt{ri-align count} command consists of a series of lines with each one containing the read name; a fraction with the length of the longest suffix of the read that occurs in the database as the numerator, and the length of the read as the denominator; and the number of times it occurs.  The fields are separated by tabs.  For the example in the previous section, the output is as follows:
\begin{verbatim}
$   ri-align count dengue data/dengue/reads.fq
>   simulated.0  100/100  22
>   simulated.1  100/100  2
>   simulated.2  100/100  2
>   simulated.3  100/100  2
...
\end{verbatim}
With some errors in the reads, the output changes as follows:
\begin{verbatim}
$   ri-align count dengue data/dengue/reads_w_errors.fq
>   simulated.0.3edits  44/100  492
>   simulated.1.2edits  78/100  1
>   simulated.2.1edits  81/100  2
>   simulated.3.0edits  100/100 2
...
\end{verbatim}
This means, for example, that after we add errors to {\tt simulated.0} to create {\tt simlulated.0.3edits}\,, the last 44 characters of the edited read occur as a substring 492 times in the the database, but the last 45 characters never occur as a substring. 

\section{Interpreting Locations}
\label{sec:locations}

The output of the \texttt{ri-align locate} is in the SAM format~\cite{LHW+09}. The SAM format begins with a line specifying the name and length of each sequence in the database (2042 for dengue), and follows with a line for each read consisting of the following tab-separated fields: read name, flag, reference name, position, MAPQ score, CIGAR string, reference name for the next read in the read-pair, position of the next read in the read-pair, observed template length, read sequence, read quality scores, and miscellaneous tags.

The most important fields are the reference name, which identifies the sequence in the database containing the occurrence, and the position, which specifies the location of the occurrence of the read is in that sequence.  Since the $r$-index does not currently support mapping-quality calculation or read-pair/mate-pair alignment, the MAPQ field, the read-pair fields and the observed template length field will all contain their respective ``unknown''-type values.  (For more details on the SAM format, see
\linebreak
{\tt https://samtools.github.io/hts-specs/SAMv1.pdf}\,.)

For our running example without errors, the output (limited to one occurrence per read) is
\begin{verbatim}
$   ri-align --max-hits 1 locate dengue data/dengue/reads.fq
>   @HD     VN:1.6  SO:unknown
>   @SQ     SN:gb:KY474305|Organism:Dengue  LN:10676
>   @SQ     SN:gb:JN638344|Organism:Dengue  LN:10735
...
>   @SQ     SN:gb:AY835999|Organism:Dengue  LN:10727
>   simulated.0	0 gb:GQ868562|Organism:Dengue	2525	255	100M	*	0	0
    TGG...CTA	~~~...~~~	NH:i:22
>   simulated.1	0	gb:KY474306|Organism:Dengue	10001	255	100M	*	0	0
    CAT...AGC	~~~...~~~	NH:i:2
>   simulated.2	0	gb:KY474306|Organism:Dengue	8833	255	100M	*	0	0
    GAT...TTG	~~~...~~~	NH:i:2
>   simulated.3	0	gb:KY474306|Organism:Dengue	8149	255	100M	*	0	0
    CTA...GAA	~~~...~~~	NH:i:2
...
\end{verbatim}
with ellipses shortening the 100-character reads and strings of \verb!~! symbols.

At the moment, {\tt ri-align locate} reports only on reads that match perfectly; of course, users have the option to use {\tt ri-align count} to find the lengths of the longest suffixes that occur in the database, then search with {\tt ri-align locate} only for those suffixes.  This can be used to generate seeds for approximate pattern matching of reads with errors or unseen variations.  Since Illumina reads tend to have more errors towards the end, users may want to truncate their reads to obtain longer matching suffixes.

\section{Acknowledgements}
\label{sec:acks}

AK and CB were funded by the NIH though NIAID grant R01AI141810-01 and by the NSF through grant IIS-1618814.  TM and BL were funded by the NIH through NIGMS grant R01GM118568 and by the NSF through IIS grant 1349906.  TG was funded by Fondecyt grant 1171058 and a start-up grant from Dalhousie University.  GM was partially funded by PRIN grant 201534HNXC and by INdAM-GNCS Project 2019 ``Innovative methods for the solution of medical and biological big data''.

\end{document}